\documentclass[11pt,twoside]{article}

\usepackage{amsmath}
\usepackage{amssymb}
\usepackage{cite}
\usepackage{ccaption}
\usepackage{hyperref}
\usepackage{graphicx}
\captiondelim{.}

\begin{document}
\newcommand{\pst}{\hspace*{1.5em}}

\newcommand{\rigmark}{\em Journal of Russian Laser Research}
\newcommand{\lemark}{\em Volume 30, Number 5, 2009}

\newcommand{\be}{\begin{equation}}
\newcommand{\ee}{\end{equation}}
\newcommand{\bm}{\boldmath}
\newcommand{\ds}{\displaystyle}
\newcommand{\bea}{\begin{eqnarray}}
\newcommand{\eea}{\end{eqnarray}}
\newcommand{\ba}{\begin{array}}
\newcommand{\ea}{\end{array}}
\newcommand{\arcsinh}{\mathop{\rm arcsinh}\nolimits}
\newcommand{\arctanh}{\mathop{\rm arctanh}\nolimits}
\newcommand{\bc}{\begin{center}}
\newcommand{\ec}{\end{center}}

\thispagestyle{plain}

\label{sh}


\begin{center} {\Large \bf
\begin{tabular}{c}
Contextuality in tree-like graphs
\end{tabular}
 } \end{center}

\bigskip

\bigskip

\begin{center} {\bf
A.A. Strakhov$^1$ and V.I. Man'ko$^2$
}\end{center}

\medskip

\begin{center}
{\it
$^1$Moscow Institute of Physics and Technology \\
Institutskii per. 9, Dolgoprudnyi, Moscow Region 141700, Russia

\smallskip

$^2$P.N. Lebedev Physical Institute, Russian Academy of Sciences\\
Leninskii Prospect 53, Moscow 119991, Russia
}
\smallskip

\end{center}


\medskip
\begin{abstract}
The contextuality problem connected with existence of joint probability
distribution creating all the given marginals is studied. It is shown for several examples considered previously in literature that there exist some new solutions for the joint probability distributions providing the given marginals and numerical example of the new solution is demonstrated.
\end{abstract}
\section{Introduction}
\pst
Quantum correlations in composite systems provide the important resource for quantum technologies like quantum computing, quantum communications, quantum teleportation, etc~\cite{1}. The entanglement~\cite{2} is one of the subsystems correlations phenomenon. The other kind of quantum correlations is the existence of quantum discord~\cite{3} also associated with the composite system structure. The third kind of quantum correlations can be considered on example of contextuality phenomenon~\cite{4,5,6,7}.

Recently tomographic approach to quantum state description was suggested~\cite{8,9}. In this approach the states are identified with fair probability distributions. For qudit systems these distributions are associated with finite probability vectors. The problem of relationship between contextuality and tomographic probability vectors was discussed in~\cite{10}. There exist the problem of constructing a joint probability distribution if the marginal distributions are known~\cite{11}. In quantum aspects the problem was studied recently in~\cite{12}.

The aim of our work is to consider the relationship between marginal distributions and joint probability distributions due to possibility of describing the quantum states by quantum-tomographic probability vectors. We focus on studying the existence of solutions for particular linear equations connecting vectors corresponding to joint probability distributions of composite system with the marginal probability vectors. We will show that there exists a set of solutions in comparison to one solution discussed in~\cite{6}.

\section {A simple formula for joint probability distribution}
\pst
In~\cite{6} it was shown that for complementarity graphs without cycles we can construct joint probability distribution out of pairwise joint distributions for mutual random variables. The principle of construction is the following: the product of probability distributions corresponding to the edges of the graph (denoted by the set $E(G)$) is divided by the product of probabilities of vertices, connected with 2 or more others, i.e., a vertex $i\in V(G)$(the set of vertices) of degree $d(i)$ (the number of neighboring vertices) appears $d(i)-1$ times in the denominator.
\be
p(A_1,A_2,...,A_n)=\frac {\prod_{(i,j)\in E(G)}p(A_i,A_j)}{\prod_{i\in V(G)}p(A_i)^{d(i)-1}}.
\label{eq1}
\ee
For instance, let's take two graphs on Figure 1. For them the joint distributions are:
\begin{figure}[ht]
\bc \includegraphics[width=8cm]{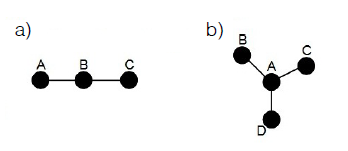}
\ec
\vspace{-4mm}
\caption{}
\end{figure}

\be
\text{a)}\;p(A,B,C)=\frac {p(A,B)p(B,C)}{p(B)};\quad\text{b)}\;p(A,B,C,D)=\frac {p(A,B)p(A,C)p(A,D)}{p(A)^2}.
\label{eq2}
\ee
\pst
In this paper we will discuss two types of graphs without cycles - linear chains and intersections - vertices with degree bigger than 1. Out of them as basic parts one will be able to analyze arbitrary graphs without cycles. We will show, that for such graphs there are more than one described by (\ref{eq1}) probability distributions.

\section {Contextuality in the shortest chain}
\pst
Let us discuss the case of the simplest chain, pictured on Figure 1(a) - with 3 elements and each of them is a dichotomic random variable. We will show that in general we are able to construct a set of joint probability distributions out of given pairwise distributions. So, we start with two distributions $\vec{P}_{ab}$ and $\vec{P}_{bc}$:
$$\vec{P}=\vec{P}_{ab}=\begin{pmatrix}
P(a=0,b=0) \\
P(a=0,b=1) \\
P(a=1,b=0) \\
P(a=1,b=1)
\end{pmatrix};
\quad
\vec{\Pi}=\vec{P}_{bc}=\begin{pmatrix}
P(b=0,c=0) \\
P(b=0,c=1) \\
P(b=1,c=0) \\
P(b=1,c=1)
\end{pmatrix}.$$
Probability theory provides the obvious constraints on these distributions:
\be
\begin{cases}
P_1+P_3=\Pi_1+\Pi_2 \\
\sum\limits_{i=1}^{4} P_i=1 \\
\sum\limits_{i=1}^{4} \Pi_i=1 \\
\end{cases}.
\label{eq3}
\ee
Now let's suppose the existence of joint probability distribution for all random variables $\vec{X}=\vec{P}_{abc}$, which also satisfies normalization constraint:
\be
\sum\limits_{i=1}^{8} X_i=1.
\label{eq4}
\ee
If the distribution $\vec{X}$ exists, then we will be able to calculate its marginals $\vec{P}$ and $\vec{\Pi}$. This fact and constraints (\ref{eq3}) and (\ref{eq4}) provide a set of equations on $\vec{X}$:
\be
\begin{cases}
P_1+P_3=\Pi_1+\Pi_2 \\
P_2+P_4=\Pi_3+\Pi_4 \\
X_1+X_2=P_1 \\
X_3+X_4=P_2 \\
X_5+X_6=P_3 \\
X_7+X_8=P_4 \\
X_1+X_5=\Pi_1 \\
X_2+X_6=\Pi_2 \\
X_3+X_7=\Pi_3 \\
X_4+X_8=\Pi_4 \\
\sum\limits_{i=1}^{4} P_i=1 \\
\sum\limits_{i=1}^{4} \Pi_i=1 \\
\sum\limits_{i=1}^{8} X_i=1 \\
\end{cases}
\Rightarrow
\begin{cases}
P_1+P_3=\Pi_1+\Pi_2 \\
X_2=P_1-X_1 \\
X_4=P_2-X_3 \\
X_5=\Pi_1-X_1 \\
X_6=\Pi_2-P_1+X_1=P_3-\Pi_1+X_1 \\
X_7=\Pi_3-X_3 \\
X_8=\Pi_4-P_2+X_3=P_4-\Pi_3+X_3 \\
\end{cases}.
\label{eq5}
\ee
As we can see, distribution $\vec{X}$ is determined up to free parameters $X_1$ and $X_3$. But actually, they are not arbitrary numbers. We must also remember about nonnegativity of $\vec{P}$, $\vec{\Pi}$ and $\vec{X}$:
\be
\forall i:
\begin{cases}
P_i\ge 0 \\
\Pi_i\ge 0 \\
X_i\ge 0 \\
\end{cases}.
\label{eq6}
\ee
Considering together (\ref{eq5}) and (\ref{eq6}), we achieve the following constraints on two independent parameters $X_1$ and $X_3$:
\be
\begin{cases}
X_1\leq P_1 \\
X_1\leq \Pi_1 \\
X_1\geq P_1-\Pi_2 \\
X_1\geq \Pi_1-P_3 \\
\end{cases};
\quad
\begin{cases}
X_3\leq P_2 \\
X_3\leq \Pi_3 \\
X_3\geq P_2-\Pi_4 \\
X_3\geq \Pi_3-P_4 \\
\end{cases}.
\label{eq7}
\ee
From (\ref{eq7}) it is obvious, that for arbitrary probability distributions $\vec{P}$ and $\vec{\Pi}$ there always exist closed intervals for parameters $X_1$ and $X_3$, which determine the whole distribution $\vec{P}_{abc}=\vec{X}$ out of (\ref{eq5}). So the system is noncontextual and in general case there is a set of joint probability distributions. In certain cases, when the lengths of these intervals are equal to 0, the joint distribution $\vec{P}_{abc}$ becomes uniquely determined and then it is generated by (\ref{eq2}a).

\section {Contextuality in chain of arbitrary length}
\begin{figure}[ht]
\bc \includegraphics[width=6cm]{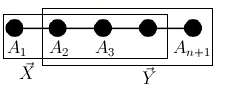}
\ec
\vspace{-4mm}
\caption{}
\end{figure}
\pst
Now, let's move to the case of the arbitrary length chain of dichotomic random variables. As in previous case, it also can be shown, that the set of joint probability distributions $\vec{P}_{A_1,A_2,...,A_{n+1}}=\vec{Z}$ always exists. We will do it, using mathematical induction.
Let's denote the amount of vertices of the chain by Q. The base case is the chain with $Q=3$, that was already considered in previous section. The inductive step is a transition from $Q=N$ to $Q=N+1$. So, we suppose that joint probabilities $\vec{X}$ and $\vec{Y}$ (Figure 2) for two subchains with $Q=N$ exist. Binary representations of indices of $\vec{X}$ and $\vec{Y}$ fit elementary outcomes of dichotomic random variables. Then, using arguments and constraints similar to those that were in previous section, we will achieve the following equations:
$$
\begin{cases}
X_1+X_{2^{n-1}+1}=Y_1+Y_2 \\
... \\
X_{2^{n-1}-1}+X_{2^{n}-1}=Y_{2^{n}-3}+Y_{2^{n}-2} \\
X_{2^{n-1}}+X_{2^{n}}=Y_{2^{n}-1}+Y_{2^{n}} \\
Z_1+Z_2=X_1 \\
Z_3+Z_4=X_2 \\
... \\
Z_{2^{n+1}-3}+Z_{2^{n+1}-2}=X_{2^{n}-1} \\
Z_1+Z_{2^{n}+1}=Y_1 \\
Z_3+Z_{2^{n}+3}=Y_3 \\
... \\
Z_{2^{n}-1}+Z_{2^{n+1}-1}=Y_{2^{n}-1} \\
Z_{2^{n+1}}=1-\sum\limits_{i=1}^{2^{n+1}-1}Z_i=Z_{2^{n}-1}+X_{2^{n}}-Y_{2^{n}-1}
\end{cases}
\Rightarrow
\begin{cases}
Z_2=X_1-Z_1 \\
Z_{2^{n}+1}=Y_1-Z_1 \\
... \\
Z_{2^{n}}=X_{2^{n-1}}-Z_{2^{n}-1} \\
Z_{2^{n+1}-1}=Y_{2^{n}-1}-Z_{2^{n}-1} \\
Z_{2^{n}+2}=X_{2^{n-1}+1}-Z_{2^{n}+1}=Z_1+X_{2^{n-1}+1}-Y_1 \\
... \\
Z_{2^{n+1}-2}=Z_{2^{n}-3}+X_{2^{n}-1}-Y_{2^{n}-3} \\
\end{cases}.
$$
The whole distribution $\vec{Z}$ is determined by $2^{n-1}$ parameters. As in previous section, using nonnegativity of probabilities $\vec{X}$, $\vec{Y}$ and $\vec{Z}$ we will achieve the following constraints on parameters $Z_1$, $Z_3$,..., $Z_{2^{n-1}}$:
\be
\begin{cases}
Z_1\leq X_1 \\
Z_1\leq Y_1 \\
Z_3\leq X_2 \\
Z_3\leq Y_2 \\
... \\
Z_{2^{n}-1}\leq X_{2^{n-1}} \\
Z_{2^{n}-1}\leq Y_{2^{n-1}} \\
Z_1\geq X_{1}-Y_{2} \\
Z_1\geq Y_{1}-X_{2^{n-1}+1} \\
... \\
Z_{2^{n}-3}\geq X_{2^{n-1}-1}-Y_{2^{n}-2} \\
Z_{2^{n}-3}\geq Y_{2^{n}-3}-X_{2^{n}-1} \\
Z_{2^{n}-1}\geq X_{2^{n-1}}-Y_{2^{n}} \\
Z_{2^{n}-1}\geq Y_{2^{n}-1}-X_{2^{n}} \\
\end{cases}.
\label{eq8}
\ee
Now it's obvious, that parameters $Z_1$, $Z_3$,..., $Z_{2^{n-1}}$ taken from the closed $2^{n-1}$-dimensional parallelepiped defined by (\ref{eq8}) generate a set of joint probability distributions $\vec{P}_{A_1,A_2,...,A_{n+1}}=\vec{Z}$. In certain cases, if this parallelepiped is a single dot, the uniquely determined probability distribution $\vec{P}_{A_1,A_2,...,A_{n+1}}$ is equal to one defined by (\ref{eq1}).

\section{Contextuality in intersections with degree bigger than 1}
\pst
\begin{figure}[ht]
\bc \includegraphics[width=5cm]{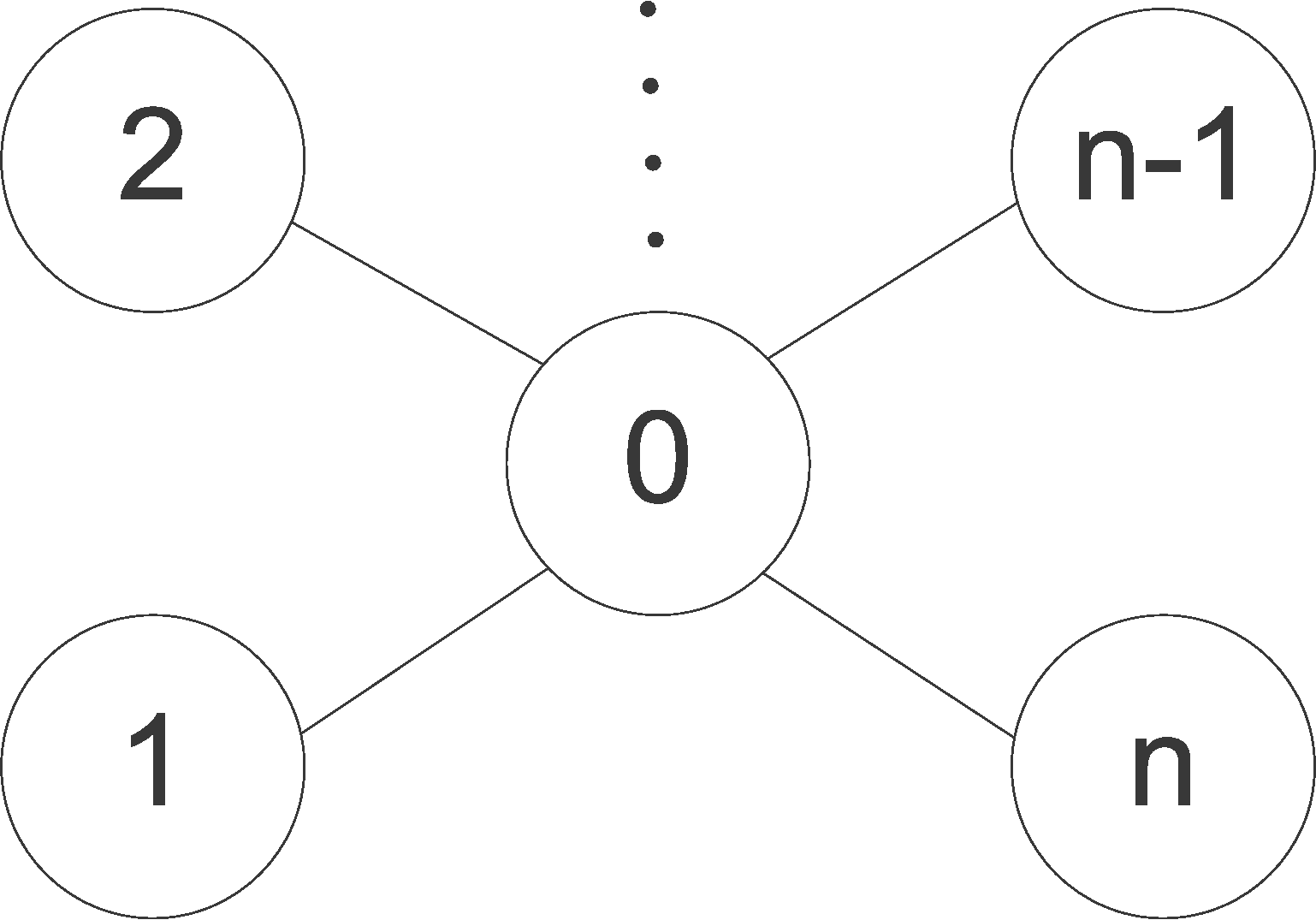}
\ec
\vspace{-4mm}
\caption{}
\end{figure}

Now let's describe graphs with one central vertex with arbitrary degree and other vertices having degree equal to 1 (Figure 3). The way of studying them is the same as for chains - inductive. The base of induction is a minimal chain with 3 vertices. It has been already studied in the previous sections. The inductive step is an addition of a new vertex, connected only with central.

The designations are the following: the central vertex is denoted by 0 and the boundary vertices are denoted by numbers from 1 to n, so we have the graph with n+1 vertices. Then the new joint probability distribution for this graph must provide all marginals - joint probabilities for all subgraphs with n vertices and less. For the proof of existence of joint distribution for n+1 vertices it will be enough to use only two marginals - for instance, for subgraphs without vertex 1 and without vertex n respectively. We use the following designations: the joint probability distribution for the whole graph is $\vec{P}_{012...n}=\vec{Y}$, the marginals for subgraphs without vertex 1 and vertex n are $\vec{P}_{02...n}=\vec{X}^1$ and $\vec{P}_{012...n-1}=\vec{X}^n$ respectively. As in previous sections, binary representations of indices of probability vectors fit elementary outcomes of dichotomic random variables. Now, we are able to write equations connecting $\vec{Y}$, $\vec{X}^1$ and $\vec{X}^n$:
$$\begin{cases}
X^1_1+X^1_2=X^n_1+X^n_{1+2^{n-2}} \\
X^1_3+X^1_4=X^n_2+X^n_{2+2^{n-2}} \\
... \\
X^1_{2^n-1}+X^1_{2^n}=X^n_{2^n-2^{n-2}}+X^n_{2^n} \\
Y_1+Y_2=X^n_1 \\
Y_3+Y_4=X^n_2 \\
... \\
Y_{2^{n+1}-1}+Y_{2^{n+1}}=X^n_{2^n} \\
Y_1+Y_{1+2^{n-1}}=X^1_1 \\
Y_2+Y_{2+2^{n-1}}=X^1_2 \\
... \\
Y_{2^{n+1}-2^{n-1}}+Y_{2^{n+1}}=X^1_{2^n} \\
\sum\limits_{i=1}^{2^n} X^1_i=\sum\limits_{i=1}^{2^n} X^n_i=\sum\limits_{i=1}^{2^{n+1}} Y_i=1 \\
\end{cases}
\Rightarrow$$
$$
\Rightarrow
\begin{cases}
Y_2=X^n_1-Y_1 \\
Y_{1+2^{n-1}}=X^1_1-Y_1=X^n_1-X^1_2+X^n_{1+2^{n-2}}-Y_1 \\
Y_{2+2^{n-1}}=X^1_2-Y_2=X^1_2-X^n_1+Y_1 \\
... \\
Y_{2^{n-1}}=X^n_{2^{n-2}}-Y_{2^{n-1}-1} \\
Y_{2^n-1}=X^1_{2^{n-1}-1}-Y_{2^{n-1}-1}=X^n_{2^{n-2}}-X^1_{2^{n-1}}+X^n_{2^{n-1}}-Y_{2^{n-1}-1} \\
Y_{2^n}=X^1_{2^{n-1}}-Y_{2^{n-1}}=X^1_{2^{n-1}}-X^n_{2^{n-2}}+Y_{2^{n-1}-1} \\
Y_{2^n+2}=X^n_{2^{n-1}+1}-Y_{2^n+1} \\
Y_{2^n+2^{n-1}+1}=X^1_{2^{n-1}+1}-Y_{2^n+1}=X^n_{2^{n-1}+1}-X^1_{2^{n-1}+2}+X^n_{2^{n-1}+2^{n-2}+1}-Y_{2^n+1}\\
Y_{2^n+2^{n-1}+2}=X^1_{2^{n-1}+2}-Y_{2^n+2}=X^1_{2^{n-1}+2}-X^n_{2^{n-1}+1}+Y_{2^n+1} \\
... \\
Y_{2^n+2^{n-1}}=X^n_{2^{n-1}+2^{n-2}}-Y_{2^n+2^{n-1}-1} \\
Y_{2^{n+1}-1}=X^1_{2^n-1}-Y_{2^n+2^{n-1}-1}=X^n_{2^{n-1}+2^{n-2}}-X^1_{2^n}+X^n_{2^n}-Y_{2^n+2^{n-1}-1}\\
Y_{2^{n+1}}=X^1_{2^n}-Y_{2^n+2^{n-1}}=X^1_{2^n}-X^n_{2^{n-1}+2^{n-2}}+Y_{2^n+2^{n-1}-1} \\
\end{cases}.$$
As one can see, the distribution $\vec{Y}$ is determined by parameters $Y_1$, $Y_3$,..., $Y_{2^{n-1}-1}$, $Y_{2^n+1}$, $Y_{2^n+3}$,..., $Y_{2^n+2^{n-1}-1}$. Now, we must use the nonnegativity of probability distributions $\vec{Y}$, $\vec{X}^1$ and $\vec{X}^n$ and we will get the following constraints on these parameters:
\be
\begin{cases}
X^n_1-X^1_2\leq Y_1\leq X^1_1=X^n_1-X^1_2+X^n_{1+2^{n-2}} \\
0\leq Y_1\leq X^n_1 \\
... \\
X^n_{2^{n-2}}-X^1_{2^{n-1}}\leq Y_{2^{n-1}-1}\leq X^1_{2^{n-1}-1}=X^n_{2^{n-2}}-X^1_{2^{n-1}}+X^n_{2^{n-1}} \\
0\leq Y_{2^{n-1}-1}\leq X^n_{2^{n-2}} \\
X^n_{2^{n-1}+1}-X^1_{2^{n-1}+2}\leq Y_{2^n+1}\leq X^1_{2^{n-1}+1}=X^n_{2^{n-1}+1}-X^1_{2^{n-1}+2}+X^n_{2^{n-1}+2^{n-2}+1} \\
0\leq Y_{2^n+1}\leq X^n_{2^{n-1}+1} \\
... \\
X^n_{2^{n-1}+2^{n-2}}-X^1_{2^n}\leq Y_{2^n+2^{n-1}-1}\leq X^1_{2^n-1}=X^n_{2^{n-1}+2^{n-2}}-X^1_{2^n}+X^n_{2^n} \\
0\leq Y_{2^n+2^{n-1}-1}\leq X^n_{2^{n-1}+2^{n-2}} \\
\end{cases}.
\label{eq9}
\ee
From (\ref{eq9}) it becomes obvious, that for every $\vec{X}^1$ and $\vec{X}^n$ there exists a set of joint probability distributions $\vec{P}_{012...n}=\vec{Y}$ yielding proper pairwise joint probability distributions for all edges of graph.

For instance, in case of $n=3$, which is presented on the Figure 1(b) (and, in new designations, on the Figure 4) the set of equations is the following:
\be
\begin{cases}
X^1_1+X^1_2=X^3_1+X^3_3 \\
X^1_3+X^1_4=X^3_2+X^3_4 \\
... \\
X^1_7+X^1_8=X^n_6+X^n_8 \\
Y_1+Y_2=X^3_1 \\
Y_3+Y_4=X^3_2 \\
... \\
Y_{15}+Y_{16}=X^3_8 \\
Y_1+Y_5=X^1_1 \\
Y_2+Y_6=X^1_2 \\
... \\
Y_{12}+Y_{16}=X^1_8 \\
\sum\limits_{i=1}^8 X^1_i=\sum\limits_{i=1}^8 X^n_i=\sum\limits_{i=1}^{16} Y_i=1 \\
\end{cases}.
\label{eq10}
\ee
In this case the parameters are: $Y_1$, $Y_3$, $Y_9$ and $Y_{11}$. The constraints on them are:
\be
\begin{cases}
X^3_1-X^1_2\leq Y_1\leq X^1_1=X^3_1-X^1_2+X^3_3 \\
0\leq Y_1\leq X^3_1 \\
X^3_2-X^1_4\leq Y_3\leq X^1_3=X^3_2-X^1_4+X^3_4 \\
0\leq Y_3\leq X^3_2 \\
X^3_5-X^1_6\leq Y_9\leq X^1_5=X^3_5-X^1_6+X^3_7 \\
0\leq Y_9\leq X^3_5 \\
X^3_6-X^1_8\leq Y_{11}\leq X^1_7=X^3_6-X^1_8+X^3_8 \\
0\leq Y_{11}\leq X^3_6 \\
\end{cases}.
\label{eq11}
\ee

\section {Example of a graph with specific pairwise distributions}
\pst
In this section we will study the graph on Figure 4 in numbers using developed formalism and compare results with formulas (\ref{eq2}a) and (\ref{eq2}b).
\begin{figure}[ht]
\bc \includegraphics[width=5cm]{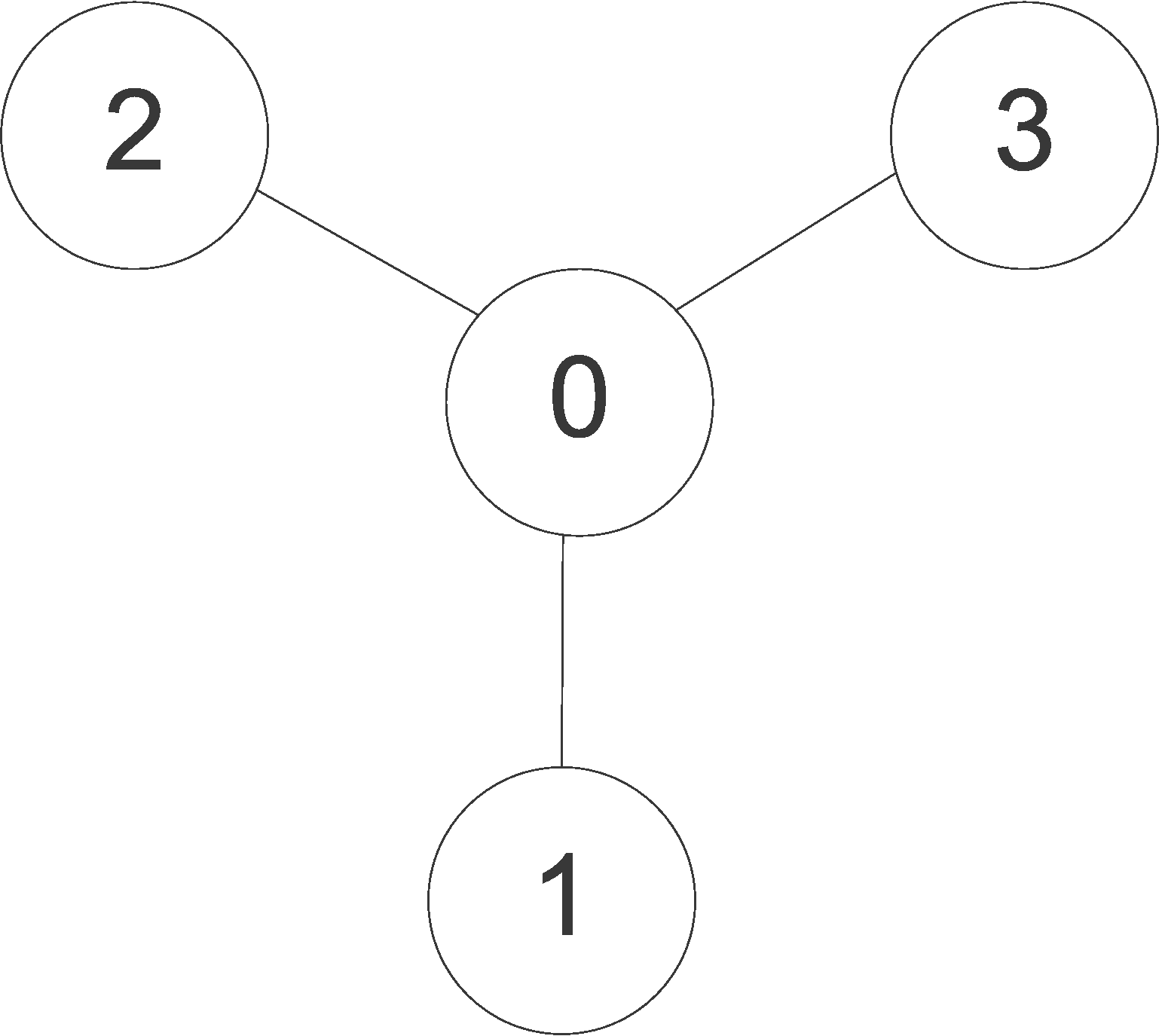}
\ec
\vspace{-4mm}
\caption{}
\end{figure}

The starting pairwise probability distributions and the distribution of central random variable are the following:
$$
\vec{P}_0=(\frac{9}{20},\;\frac{11}{20})^T;
\quad
\vec{P}_{01}=(\frac{1}{5},\;\frac{1}{4},\;\frac{2}{5},\;\frac{3}{20})^T;
$$
$$
\vec{P}_{02}=(\frac{1}{5},\;\frac{1}{4},\;\frac{7}{20},\;\frac{1}{5})^T;
\quad
\vec{P}_{03}=(\frac{1}{5},\;\frac{1}{4},\;\frac{1}{4},\;\frac{3}{10})^T.
$$
The order of indices in $\vec{P}_{01}$ ($\vec{P}_{01}$ or $\vec{P}_{10}$) has meaning in context of equations written in previous sections. For instance, to get $\vec{P}_{012}=\vec{X}^3$ we need to use results of section 3 and calculate $\vec{P}_{102}$ out of $\vec{P}_{10}$ and $\vec{P}_{02}$ and then swap 3rd component in $\vec{P}_{102}$ with 5th and 4th with 6th respectively.
Using results of section 3, we define the set of probability distributions $\vec{P}_{102}=\vec{X}^{'3}$ and $\vec{P}_{203}=\vec{X}^{'1}$, and then, $\vec{P}_{012}=\vec{X}^3$ and $\vec{P}_{023}=\vec{X}^1$:
\be
\begin{cases}
0\leq X^{'3}_1\leq \frac {1}{5} \\
\frac {1}{5}\leq X^{'3}_3\leq \frac {7}{20} \\
\end{cases}.
\label{eq12}
\ee
Following formula (\ref{eq2}a), out of $P_{10}$ and $P_{02}$ we will calculate $P_{102=000}=X^{'3}_1=\frac {4}{45}$ and $P_{102=010}=X^{'3}_3=\frac {14}{55}$. As we can see, these values are in the ranges (\ref{eq12}), so proposed in this article approach yields proper results.

Let us choose the following values out of ranges (\ref{eq12}): $X^{'3}_1=\frac {1}{10}$ and $X^{'3}_3=\frac {1}{5}$. No we are able to recover vectors $\vec{X}^{'3}$ and $\vec{X}^{'1}$, and therefore $\vec{X}^3$ and $\vec{X}^1$:
$$
\vec{X}^{'3}=(\frac {1}{10},\;\frac {1}{10},\;\frac {1}{5},\;\frac {1}{5},\;\frac {1}{10},\;\frac {3}{20},\;\frac {3}{20},\;0)^T
\Rightarrow
\vec{X}^3=(\frac {1}{10},\;\frac {1}{10},\;\frac {1}{10},\;\frac {3}{20},\;\frac {1}{5},\;\frac {1}{5},\;\frac {3}{20},\;0)^T;
$$
$$
\vec{X}^{'1}=(\frac {3}{20},\;\frac {1}{20},\;\frac {1}{10},\;\frac {1}{4},\;\frac {1}{20},\;\frac {1}{5},\;\frac {3}{20},\;\frac {1}{20})^T
\Rightarrow
\vec{X}^1=(\frac {3}{20},\;\frac {1}{20},\;\frac {1}{20},\;\frac {1}{5},\;\frac {1}{10},\;\frac {1}{4},\;\frac {3}{20},\;\frac {1}{20})^T.
$$
Now, using inequalities (\ref{eq11}), we are able to obtain constraints on parameters $Y_1$, $Y_3$, $Y_9$ and $Y_{11}$ of the joint probability distribution $\vec{P}_{0123}=\vec{Y}$:
\be
\begin{cases}
\frac {1}{20}\leq Y_1 \leq \frac{1}{10} \\
0 \leq Y_3 \leq \frac{1}{20} \\
0 \leq Y_9 \leq \frac{1}{10} \\
\frac {3}{20}\leq Y_{11} \leq \frac{3}{10} \Leftrightarrow Y_{11}=\frac {3}{20}\\
\end{cases}.
\label{eq13}
\ee
Let us choose the following values from (\ref{eq13}): $Y_1=\frac {1}{20}$, $Y_3=0$, $Y_9=\frac {1}{20}$ and $Y_{11}=\frac {3}{20}$. Now, out of (\ref{eq10}) we are able to obtain the whole distribution $\vec{Y}=\vec{P}_{0123}$:
\be
\vec{P}_{0123}=(\frac {1}{20},\;\frac {1}{20},\;0,\;\frac {1}{10},\;\frac {1}{10},\;0,\;\frac {1}{20},\;\frac {1}{10},\;\frac {1}{20},\;\frac {3}{20},\;\frac {3}{20},\;\frac {1}{20},\;\frac {1}{20},\;\frac {1}{10},\;0,\;0)^T.
\label{eq14}
\ee
After some simple calculations one can confirm that vector (\ref{eq14}) is really a joint probability distribution that yields marginals $\vec{P}_{01}$, $\vec{P}_{02}$ and $\vec{P}_{03}$. Let's check the first component $\vec{P}_{0123}$ calculated using (\ref{eq2}b):
$$
P(A=0,B=0,C=0,D=0)=\frac {p(A=0,B=0)p(A=0,C=0)p(A=0,D=0)}{p(A=0)^2}=\frac {\frac {1}{5}\cdot\frac {1}{5}\cdot\frac {1}{5}}{\frac {9}{20}}=\frac {4}{225}.
$$
As one can see, it differs from $Y_1=P_{0123=0000}=\frac {1}{20}$. The set of probability distributions defined by ranges (\ref{eq13}) doesn't include the distribution defined by (\ref{eq2}b) because we chose marginals $\vec{P}_{012}$ and $\vec{P}_{023}$ different from that are defined by (\ref{eq2}a). Surely, if we choose them, new ranges (\ref{eq13}) will include the distribution defined by (\ref{eq2}b).

\section {Conclusions}
\pst
To resume we point out main results of our work. For the system of dichotomic random variables with specific pairwise probability distributions which structure is described by complementarity graph without circles there exist extra to already known solutions for the joint probability distribution. The proof of this point can be easily generalized to the case of discrete random variables with more than 2 elementary outcomes. We shall apply these results to qudit tomograms in future publications.

\end{document}